\documentclass[prd,aps,floats,twocolumn,twoside,preprintnumbers,
superscriptaddress,floatfix,nofootinbib,showpacs]{revtex4}
\usepackage{graphicx,pstricks,rotating,amsmath}
\usepackage{slashed}
\usepackage{hyperref}

\begin{document}

\title{Photon-neutrino interaction in $\theta$-exact
covariant noncommutative field theory}

\author{R. Horvat}
\affiliation{Institute Rudjer Bo\v{s}kovi\'{c}, Bijeni\v{c}ka 54 10000 Zagreb, Croatia}
\author{D. Kekez}
\affiliation{Institute Rudjer Bo\v{s}kovi\'{c}, Bijeni\v{c}ka 54 10000 Zagreb, Croatia}
\author{P. Schupp}
\affiliation{Jacobs University Bremen,
	Center for Mathematics, Modeling and Computing, Campus Ring 1,
	28759 Bremen, Germany}
\author{J. Trampeti\'{c}}
\affiliation{Institute Rudjer Bo\v{s}kovi\'{c}, Bijeni\v{c}ka 54 10000 Zagreb, Croatia}
\affiliation{Max-Planck-Institut f\"ur Physik,
	(Werner-Heisenberg-Institut),
  	 F\"ohringer Ring 6, D-80805 M\"unchen, Germany}
\author{J. You}
\affiliation{Institute Rudjer Bo\v{s}kovi\'{c}, Bijeni\v{c}ka 54 10000 Zagreb, Croatia}
\affiliation{Mathematisches Institute G\"ottingen, Bunsenstr. 3-5,
	37073 G\"ottingen, Germany}


\begin{abstract}Photon-neutrino interactions arise quite naturally in noncommutative field theories.
Such couplings are absent in ordinary field theory and imply experimental lower bounds
on the energy scale $\Lambda_{\rm NC} \sim |\theta|^{-2}$ of noncommutativity.
Using non-perturbative methods and a Seiberg-Witten map based covariant approach
to noncommutative gauge theory, we obtain $\theta$-exact expressions for the interactions,
thereby eliminating previous restrictions to low-energy phenomena.
We discuss implications for plasmon decay, neutrino charge radii, big bang nucleosynthesis and ultrahigh energy cosmic rays.
Our results behave reasonably throughout all interaction energy scales,
thus facilitating further phenomenological applications.
\end{abstract}



\maketitle

\section{Introduction}

Neutrinos, being neutral particles, cannot couple directly to photons.
They can couple indirectly via weak interaction and other effects, potentially
leading to significant observational effects in astrophysics~\cite{Adams-1963,Bandyopadhyay:1969ck}.
The photon-neutrino scattering cross section predicted by the Standard Model is, however,
exceedingly small and likely of little practical importance in astrophysics, 
see e.g.~\cite{Dicus:1993iy}.
Field theories on noncommutative (NC) spaces offer a different interaction channel 
for neutrinos and photons:
Such theories have been introduced as effective
models\footnote{These models should be understood as effective theories
and are not necessarily renormalizable.}
for the quantum geometric structure of spacetime
that is generically expected in any reasonable quantum theory of gravity
(including string theory). The presence of gauge fields influences such
noncommutative structures and neutrinos propagating in
the modified background feel the effect. The non-zero star commutator
\begin{equation}
[ A_\mu\stackrel{\star}{,}\Psi]= A_\mu\star\Psi-\Psi\star  A_\mu.
\label{1.1}
\end{equation}
is a striking illustration of this effect in noncommutative $U_{\star}(1)$ theory: 
Fermions
that are neutral or even sterile in the commutative limit can nevertheless 
directly couple to a $U_{\star}(1)$
gauge field. In this article we shall investigate such noncommutativity induced 
interactions using methods that are non-perturbative in the noncommutativity 
parameters~\cite{Schupp:2008fs}.

Studies in noncommutative particle
phenomenology \cite{Hinchliffe:2002km}, aiming at predicting possible
experimental signatures and estimating bounds on space-time
noncommutativity from existing experimental data, started around the time that
string theory indicated that noncommutative gauge theory could
be one of its low-energy effective
theories \cite{Seiberg:1999vs}. Early attempts with simple models based on star products
(with expressions like the one given above) soon ran into several serious difficulties:
(i)~Such theories have no local gauge invariant quantities,
(ii)~fields like $ A_\mu$ do not
transform covariantly under coordinate changes~\cite{Jackiw:2001jb} and
(iii)~there are unphysical restrictions on representations and charges.
These shortcomings are overcome in an approach based on Seiberg-Witten (SW) maps,
which enables one to deform commutative gauge theories with essentially
arbitrary gauge group and
representation~\cite{Madore:2000en,Jurco:2000ja,Bichl:2001gu,Jurco:2000fb,Jurco:2001my,Jurco:2001rq}.
Since this approach also fixes problems of non-covariance under coordinate changes,
we shall refer to this class of theories as \emph{covariant noncommutative field theories}.%
\footnote{All these models are consistent, contrary to slightly misleading statements in
a recently published ``no-go theorem'' \cite{Chaichian:2009uw},
where a mal-constructed model is shown to lead to contradictions.}
After some initial enthusiasm, renormalizability of these theories turned out
to be a delicate issue. In \cite{Wulkenhaar:2001sq,Grimstrup:2002af} a Dirac fermion 4-vertex was
identified as one of the major culprits in this issue. This vertex is absent
in theories with chiral fermions \cite{Buric:2007ix,Martin:2009sg}. 
Particularly well behaved are theories where
the gauge fields couple via a star commutator (\ref{1.1}) to fermions,
as is the case in our model: Cancelations between fermion and boson loops lead
to softened UV/IR coupling \cite{Matusis:2000jf}. Regardless of the question of
renormalizability, SW-map based models serve as effective theories for some
of the quantum geometric effects expected in more fundamental theories such
as string theory and quantum gravity.
In the Seiberg-Witten map approach,  noncommutative fields $ A_\mu$, $ \Psi$,\ldots and
gauge transformation parameters $ \Lambda$ are interpreted as non-local, enveloping algebra-valued
functions of their commutative counterparts $a_\mu$, $\psi$, $\lambda$ and of the
noncommutative parameters $\theta^{\mu\nu}$, in such a way that
ordinary gauge transformations of the commutative fields induce noncommutative
gauge transformations of the noncommutative fields.
This procedure allows the construction of noncommutative extensions of
important particle physics models like the NC Standard Model (NCSM) and GUT models
\cite{Bichl:2001cq,Calmet:2001na,Behr:2002wx,Aschieri:2002mc,Melic:2005fm,Melic:2005am,Martin:2011un},
as well as various follow-on studies with NC modifications of particle physics
\cite{Ohl:2004tn,Ohl:2004ke,Melic:2005su,Alboteanu:2005gj,
Alboteanu:2006hh,Alboteanu:2007bp,Alboteanu:2007by,Buric:2007qx}.
In more recent development, it has been found that SW expanded
models at first order in~$\theta$ are well-behaved regarding
anomalies and renormalizability. For example the  NCSM \cite{Calmet:2001na} at
$\theta$-order appears to be anomaly free \cite{Martin:2002nr,Brandt:2003fx},
has remarkably well-behaved one-loop quantum corrections
\cite{Buric:2006wm} and breaks Lorentz symmetry; see also
\cite{Bichl:2001cq,Latas:2007eu,Martin:2009vg,Tamarit:2009iy,Buric:2010wd}.

To avoid strong backgrounds from known processes, considerable
efforts in noncommutative phenomenology has been directed at
interactions which are suppressed in Standard Model settings.
One important candidate or this type is the aforementioned tree-level coupling of neutrinos
with photons or, more precisely, plasmons.
Such interactions have already been studied in the framework of
noncommutative gauge theories defined by Seiberg-Witten
maps \cite{Schupp:2002up, Minkowski:2003jg}.
There, like in almost all other studies of covariant NC field theory, an expansion and cut-off
in powers of the noncommutativity parameters $\theta^{\mu\nu}$ 
was used for computational simplicity.
Such an expansion corresponds to an expansion in momenta (derivatives) and restrict the range of
validity to energies well below the noncommutativity scale  $\Lambda_{\rm NC}$.
This is usually no problem for experimental predictions because the noncommutativity parameters
$\theta^{ij}=c^{ij}/\Lambda_{\rm NC}^2$ are in general considered to be
small. There exists, however, exotic processes like ultra high energy cosmic
rays \cite{Horvat:2010sr} in which the interacting energy scale runs
higher than the current experimental
bound on the noncommutative scale $\Lambda_{\rm NC}$. Here the previously available approximate
results are inapplicable. To overcome the $\theta$-expansion and cut-off approximation,
we are using in this article
$\theta$-exact expressions and expand in powers of the coupling constant as in ordinary gauge theory.

The $\theta$-exact approach has been inspired by exact formulas for the Seiberg-Witten
map \cite{Jurco:2001my,Mehen:2000vs,Liu:2000mja,Okawa:2001mv}.
For arbitrary non-Abelian gauge theories the $\theta$-exact approach is still a challenging problem,
in particular in loop computations and at higher orders in the coupling constant.
Since perturbative renormalization and UV/IR mixing are still fairly poorly understood, 
it is not clear how to
interpret the quantum corrections and to relate them to
observations \cite{Abel:2006wj,Horvat:2010km}.
Another interesting venue for applications of
$\theta$-exact methods is the investigation of quantum corrections in
covariant noncommutative quantum field theories.
Recently it was suggested~\cite{Vilar:2009er} that noncommutative QED
might be renormalizable by adding proper counter-terms. A covariant $\theta$-exact 
version of this theory could be a very
interesting object for future studies.
First $\theta$-exact results have been published in the investigation of
UV/IR mixing in covariant NC gauge theory~\cite{Schupp:2008fs}
and later in the context of NC photon-neutrino phenomenology, namely scattering of
ultra high energy cosmic ray neutrinos on nuclei \cite{Horvat:2010sr}.
Those topics were off-limits in the old $\theta$-expansion method.

\section{Model}

In this section we recall some basic facts about covariant noncommutative gauge theory based
on Seiberg-Witten maps. We then review the derivation of $\theta$-exact interaction terms
and apply the method to the computation of neutrino-photon tri-particle vertices.
We close with a comment on an alternative covariant vertex.
Throughout the article we shall concentrate on \emph{Abelian} noncommutative gauge theory.

It is straight-forward to formulate field theories on noncommutative spaces by inserting
a star product~$\star$ between all fields in the action. This introduces ordering ambiguities
and it breaks ordinary gauge invariance (because local gauge transformations do not commute 
with star products).
In analogy to the introduction of covariant derivatives in gauge theory,
the star product can be promoted to a gauge-field dependent covariant star product~$\star'$.
Together with a gauge-field dependent covariant integral measure this leads to a noncommutative 
gauge theory.
Alternatively and in fact equivalently~\cite{Jurco:2000fb,Jurco:2001my}
it is possible to retain the original star product and instead promote all fields
to noncommutative fields and the gauge transformations to noncommutative gauge transformations.
In this construction the  ``noncommutative fields'' are obtained via
Seiberg-Witten maps~\cite{Seiberg:1999vs} and their generalizations from the original 
``commutative fields''.

With some field-ordering ``fine-tuning'', it is possible to obtain noncommutative models,
were neutrinos and other neutral fermion fields do not couple to photons
-- the minimal NC Standard Model is an example of this type. More generically, however,
electrically neutral matter fields will be promoted via (hybrid) Seiberg-Witten maps
to noncommutative fields that couple via star commutator to photons and transform in
the adjoint representation of $U_{\star}(1)$ -- this is the case for phenomenologically 
promising NC GUTs.
The inclusion of all gauge covariant coupling terms is furthermore a prerequisite 
for reasonable UV behavior.
Taking all this into account we eventually arrive at the following model of a
Seiberg-Witten type noncommutative $U_{\star}(1)$ gauge theory
\footnote{In this section we set the
coupling constant $e=1$, to restore the coupling constant one simply
substitute $a_\mu$ by $ea_\mu$, then divide the gauge-field
Lagrangian by $e^2$.} (for more details, see the discussion at the end of the section)
\begin{equation}
S=\int\left(-\frac{1}{4}F^{\mu\nu}F_{\mu\nu}+i\bar\Psi
\left(\slashed{D}-m_{\nu}\right)\Psi\right) \, d^4x
\label{swqed:action}
\end{equation}
with $F_{\mu\nu}=\partial_\mu
A_\nu-\partial_\nu A_\mu-i[A_\mu\stackrel{\star}{,}A_\nu]$ and
\begin{equation}
D_\mu\Psi=\partial_\mu\Psi-i[A_\mu\stackrel{\star}{,}\Psi] \,.
\label{swqed:actionD}
\end{equation}
All the fields in this action are images under (hybrid) Seiberg-Witten maps
of the corresponding commutative fields $a_\mu$ and $\psi$.
In the original work of Seiberg and Witten and in virtually all subsequent applications,
these maps are understood as (formal) series in powers of the noncommutativity parameter
$\theta^{\mu\nu}$. Physically, this corresponds to an expansion in momenta and is valid
only for low-energy phenomena. Here we shall not subscribe to this point of view and instead
interpret the noncommutative fields as valued in the enveloping algebra of 
the underlying gauge group.
This naturally corresponds to an expansion in powers of the gauge field $a_\mu$ 
and hence in powers of
the coupling constant. At each order in $a_\mu$ we shall determine $\theta$-exact expressions.

The expansion in powers of the commutative (gauge) field content is motivated 
from the obvious fact that in
perturbative quantum field theory one can sort the interaction
vertices by the number of external legs and this is equivalent to the
number of field operators in the corresponding interacting terms.
For any specific process and loop order there exists an upper limit on
the number of external legs. So if one expands the
noncommutative fields with respect to the formal power of the
commutative fields which are the primary fields in the theory up to
an appropriate order, the relevant vertices in a specific diagram
will automatically be exact to all orders of $\theta$.

In tree-level neutrino-photon coupling processes
 only vertices of the form $a\bar\psi\psi$ contribute, therefore an
expansion to lowest nontrivial order in $a_\mu$ (but all orders in $\theta$) is enough.
There are at least three known methods for  $\theta$-exact computations:
The closed formula derived using deformation quantization based on
Kontsevich formality maps~\cite{Jurco:2001my}, the relationship
between open Wilson lines in the commutative and noncommutative
picture~\cite{Mehen:2000vs,Okawa:2001mv}, and  direct recursive
computations using consistency conditions. For the lowest nontrivial order a
direct deduction from the recursion and consistency relations
\begin{gather}
\delta_\Lambda A_\mu \equiv \partial_\mu\Lambda +
 i[\Lambda \stackrel{\star}{,}A_\mu]
\nonumber \\= A_\mu[a_\mu+\delta_\lambda
a_\mu]-A_\mu[a_\mu] + \mathcal O(\lambda^2)\,,
\\
\delta_\Lambda\Psi \equiv i[\Lambda \stackrel{\star}{,}\Psi]
\nonumber \\
=\Psi[a_\mu+\delta_\lambda a_\mu,
\psi+\delta_\lambda\psi]
-\Psi[a_\mu,\psi] + \mathcal O(\lambda^2)\,,
\\
\Lambda[[\lambda_1,\lambda_2],a_\mu]
=[\Lambda[\lambda_1,a_\mu]\stackrel{\star}{,}
\Lambda[\lambda_2,a_\mu]]
\nonumber \\
+i\delta_{\lambda_1}\Lambda[\lambda_2,a_\mu]-i\delta_{\lambda_2}
\Lambda[\lambda_1,a_\mu]\,,
\label{SWrecurs}
\end{gather}
with the ansatz\footnote{Notation: Capital letters denote noncommutative objects,
small letters denote commutative objects, hatted capital letters denote differential
operator maps from the latter to the former.}
\begin{gather}
\Lambda=\hat\Lambda[a_\mu]\lambda=(1+\hat\Lambda^1[a_\mu]
+\hat\Lambda^2[a_\mu]+\mathcal O(a^3))\lambda\,,
\\
\Psi=\hat\Psi[a_\mu]\psi=(1+\hat\Psi^1[a_\mu]+\hat\Psi^2[a_\mu]+\mathcal
O(a^3))\psi\,,
\label{ansatz:linear}
\end{gather}
is already sufficient. Here $\hat\Psi[a_\mu]$ and $\hat\Lambda[a_\mu]$ are gauge-field
dependent differential operators that we shall now determine:
Starting
with the fermion field $\Psi$, at lowest order we have
\begin{equation}
i[\lambda\stackrel{\star}{,}\psi]=\hat\Psi[\partial\lambda]\psi\,.
\label{fermiconsiscondi}
\end{equation}
Writing the star commutator explicitly as
\begin{equation}
\begin{split}
[f\stackrel{\star}{,}g]&=f(x)(e^{i\frac{\partial_x\theta\partial_y}{2}}
-e^{-i\frac{\partial_x\theta\partial_y}{2}})g(y)\bigg|_{x=y}
\\&=2if(x)\sin(\frac{\partial_x\theta\partial_y}{2})g(y)\bigg|_{x=y}
\\&=i\theta^{ij}\bigg(\frac{\partial f(x)}{\partial x^i}\bigg)
\frac{\sin(\frac{\partial_x\theta\partial_y}{2})}
{\frac{\partial_x\theta\partial_y}{2}}
\bigg(\frac{\partial g(y)}{\partial y^{i}}\bigg)
\bigg|_{x=y}\,.
\end{split}
\label{starcommut}
\end{equation}
we observe that
\begin{equation}
\hat\Psi[a_\mu]=-\theta^{ij}a_i\star_2\partial_j
\label{fermionstar2}
\end{equation}
will fulfill the consistency relation. The generalized star product
$\star_2$~\cite{Mehen:2000vs} that appears here
is defined as follows
\begin{align}
\begin{split}
f\star_2 g&=f(x)\frac{\sin\frac{\partial_x\wedge
\partial_y}{2}}{\frac{\partial_x\wedge
\partial_y}{2}}g(y)\bigg|_{x=y}.
\end{split}
\label{star2}
\end{align}
The gauge transformation $\Lambda$ can be worked out similarly, namely
\begin{equation}
\begin{split}
0&=[\lambda_1\stackrel{\star}{,}\lambda_2]
+i\hat\Lambda[\partial\lambda_1]\lambda_2
-i\hat\Lambda[\partial\lambda_2]\lambda_1
\\
&=\frac{1}{2}([\lambda_1\stackrel{\star}{,}\lambda_2]
-[\lambda_2\stackrel{\star}{,}\lambda_1])
+i\hat\Lambda[\partial\lambda_1]\lambda_2
-i\hat\Lambda[\partial\lambda_2]\lambda_1\,,
\end{split}
\label{consisconfgt}
\end{equation}
and hence
\begin{equation}
\hat\Lambda^1=-\frac{1}{2}\theta^{ij}a_i\star_2\partial_j\,.
\label{star2gt}
\end{equation}
The gauge field $a_\mu$ requires slightly more work. The lowest order terms in its consistency
relation are
\begin{equation}
-\partial_\mu(\frac{1}{2}\theta^{ij}a_i\star_2\partial_j
\lambda)-i[\lambda\stackrel{\star}{,}a_\mu]
=A^2_\mu[a_\mu+\partial_\mu\lambda]-A^2_\mu[a_\mu]\,,
\label{conscondgf}
\end{equation}
where $A^2$ is the $a^2$ order term in the expansion of $A$ as power
series of $a$. The left hand side can be rewritten as
$-\frac{1}{2}\theta^{ij}\partial_\mu
a_i\star_2\partial_j\lambda-\frac{1}{2}
\theta^{ij}a_i\star_2\partial_\mu\partial_j\lambda
-\theta^{ij}\partial_i\lambda\star_2\partial_j a_\mu$,
where the first term comes from
$-\frac{1}{2}\theta^{ij}\partial_\mu a_i\star_2 a_j$, while the third one
comes from $-\theta^{ij}a_i\star_2\partial_j a_\mu$. After a gauge transformation,
the sum of the first and third terms equals the second term. Ultimately, we obtain
\begin{eqnarray}
A_\mu&=&a_\mu-\frac{1}{2}\theta^{ij}a_i\star_2(\partial_j
a_\mu+f_{j\mu})+\mathcal O(a^3)\,, \label{exactA}
\\
\Psi&=&\psi-\theta^{ij}a_i\star_2\partial_j\psi+\mathcal
O(a^2)\psi\,, \label{exactPsi}
\\
\Lambda&=&\lambda-\frac{1}{2}\theta^{ij}a_i\star_2\partial_j\lambda+\mathcal
O(a^2)\lambda\,, \label{exactLambda}
\end{eqnarray}
with $f_{\mu\nu}$ being the commutative field strength
$f_{\mu\nu}=\partial_\mu a_\nu-\partial_\nu a_\mu$.

Expanding the action \eqref{swqed:action} in the terms of the
commutative fields, one gets the following $\theta$-exact cubic terms up to first order in $a_\mu$:
\begin{equation}
\begin{split}
{\cal L}&=\bar\psi\gamma^\mu[a_\mu\stackrel{\star}{,}\psi]
-(\theta^{ij}a_i
\star_2 \partial_j\bar\psi)(i\slashed\partial - m_\nu)\psi\\&\quad-\bar\psi
(i\slashed\partial - m_\nu)(\theta^{ij}
a_i\star_2\partial_j\psi)+\bar\psi\mathcal{O}(a^2)\psi\,.
\label{pf:int}
\end{split}
\end{equation}
Here $\psi$ is a Dirac-type massive or massless (i.e. Weyl) neutrino field.
To extract Feynman rules in an appropriate form, we use the
arithmetic property
$i\theta^{ij}\partial_i f\star_2\partial_j g=[f\stackrel{\star}{,}g]$ to obtain
the effective neutrino-photon Lagrangian density
\begin{equation}
\begin{split}
{\cal L}
&=-(\theta^{ij}a_i\star_2\partial_j\bar\psi)(i\slashed\partial - m_\nu)\psi
\\-&\bar\psi(i\slashed\partial - m_\nu)(\theta^{ij}a_i\star_2\partial_j\psi)
\\+&i\bar\psi\gamma^{\mu}(\theta^{ij}\partial_i
a_\mu\star_2\partial_j\psi)+\mathcal O(a^2\bar\psi\psi)\,.
\label{calL}
\end{split}
\end{equation}
The rest of the derivation resembles that of  NCQED without a Seiberg-Witten
map. Just like the Moyal-Weyl star product  turns into an
exponential function in momentum space, the generalized star
product $\star_2$ turns into a function
\begin{equation}
F(q,k)=\frac{\sin \frac{q\theta k}{2}}{\frac{q\theta k}{2}},
\end{equation}
where $q$ and $k$ are the momenta of the fields involved in
the product. We notice that for tri-field interaction,
 $4$-momentum conservation $q=k-k'$ renders the function $F$ independent
of the order of momenta involved:
$F(q,k)=F(k,q)=F(k,k')=F(k',k)$.
We can hence pull out $F$ as an universal factor.
In the end we obtain the following $\theta$-exact Feynman rule for the
neutrino-photon tri-particle vertex:
\begin{eqnarray}
\Gamma^{\mu}&=& i F(q,k) \left[(\slashed k - m_{\nu})\tilde q^{\mu}
+(q\theta k)\gamma^{\mu} -\slashed q\tilde k^{\mu}\right]\,,
\label{Feynrule}
\end{eqnarray}
with the shorthand notations $q\theta k \equiv q_i \theta^{ij} k_j$ and $\tilde k^\mu = \theta^{\mu j} k_j$.

If we compare this with to the first order in $\theta$ vertex that was used in previous work, we
see that only the factor $F(q,k)$ is new. Consequently, it is this factor
that leads to modifications in $\theta$-exact
computations.

Interestingly, the first term in (\ref{Feynrule}),
\begin{equation}
\Gamma^{\mu}_\text{alt} = i F(q,k) (\slashed k - m_{\nu})\tilde
q^{\mu} \label{simplevertex} \,,
\end{equation}
is already consistent with gauge invariance on its own. This simplified vertex defines an alternative
NC theory of neutrino-photon interaction that is attractive for computations beyond tree level.
In this article we will use the full vertex which is more natural from the point of view of
NC gauge theory as it is derived from a covariant derivative.


The two choices  of NC vertices is ultimately related to a choice of generalized SW map in
the construction of the NC theory. In the remainder of
this section we shall explore this construction in more detail.
For simplicity of presentation we shall set $e=1$ and focus on the massless case.

We start with the action for a neutral massless free fermion field
\begin{equation}
S = \int \bar\psi\gamma^\mu\partial_\mu\psi \, d^4x =\int  \bar\psi\star \gamma^\mu\partial_\mu\psi\, d^4x \,,
\end{equation}
where, as indicated, a Moyal-Weyl type star product can be inserted or removed by partial integration.
Following the method of constructing a covariant NC gauge theory outlined at the beginning of this section,
we lift the factors in the action via (generalized) Seiberg-Witten maps $\hat \Psi[a_\mu]$, $\hat \Phi[a_\mu]$
to noncommutative status as follows:
\begin{equation}
S=\int  \hat\Psi(\bar\psi)\gamma^\mu \hat\Phi(\partial_\mu\psi) \, d^4x
= \int  \hat\Psi(\bar\psi)\star \gamma^\mu \hat\Phi(\partial_\mu\psi)  \, d^4x  \,.
\end{equation}
Now if the SW maps $\hat\Psi$, $\hat\Phi$ and a corresponding map $\hat\Lambda$ for the gauge parameter $\lambda$
satisfy
\begin{equation}
\begin{split}
\delta_\lambda(\hat\Psi(\bar\psi))=i[\hat\Lambda(\lambda)\stackrel{\star}{,}\hat\Psi(\bar\psi)],\\
\delta_\lambda(\hat\Phi(\partial_\mu\psi))
=i[\hat\Lambda(\lambda)\stackrel{\star}{,}\hat\Phi(\partial_\mu\psi)],
\end{split}
\end{equation}
we will have a noncommutative action that is
gauge invariant under infinitesimal commutative gauge
transformations $\delta_\lambda$ and reduces to the free fermion action in the commutative limit $\theta\to 0$.

The appropriate map $\hat\Psi$ is the one (\ref{exactPsi}) that we have already derived:
\begin{equation}
\hat\Psi(\psi)=\psi-\theta^{ij}a_i\star_2\partial_j\psi+\mathcal
O(a^2)\psi.
\end{equation}
Recalling that we are dealing with neutral fields, i.e.\ $\delta \psi = 0$ and $\delta (\partial_\mu\psi) = 0$,
we notice that we can in principle use the same map also for
$\hat\Phi$:
\begin{equation}
\hat\Phi_\text{alt}(\partial_\mu\psi)=\hat\Psi(\partial_\mu\psi)
=\partial_\mu\psi-\theta^{ij}a_i\star_2(\partial_j\partial_\mu\psi)+\mathcal
O(a^2)\psi \,.
\end{equation}
This construction is quite unusual from the point of gauge theory, as it yields a covariant derivative
term without introducing a covariant derivative. In any case the resulting action
\begin{equation}
\begin{split}
S_\text{alt}&=\int \Big(
i\bar\psi\gamma^\mu\partial_\mu\psi-i\left(\theta^{ij}\partial_j\bar\psi
\star_2 a_i\right)\gamma^\mu\partial_\mu\psi
\\+&i\bar\psi\gamma^{\mu}\left(\theta^{ij}
a_i\star_2\partial_\mu\partial_j\psi\right) \Big) \, d^4x \; + \mathcal{O}(a^2)
\end{split}
\end{equation}
is consistent and gauge invariant.
The corresponding photon-fermion interaction vertex
\begin{equation}
\Gamma^{\mu}_\text{alt}=i F(q,k) \tilde q^\mu\slashed k
\end{equation}
is surprisingly simple and therefore quite attractive for loop-level computations. The vertex satisfies
\begin{equation}
q_\mu\Gamma^{\mu}_\text{alt}=(q\cdot \tilde q)i F(q,k) \slashed k=0
\,.
\end{equation}
There is, however, a second choice for $\hat\Phi$:
\begin{equation}
\begin{split}
&\hat\Phi_{}(\partial_\mu\psi)=D^\star_\mu \hat\Psi(\psi)=\partial_\mu
\hat\Psi(\psi)-i[A_\mu\stackrel{\star}{,}\hat\Psi(\psi)]
\\
=&\partial_\mu\psi-\theta^{ij}a_i\star_2\partial_j\partial_\mu\psi+\theta^{ij}f_{i\mu}\star_2\partial_j\psi
+O(a^2)\psi \,,
\end{split}
\end{equation}
based on the well-known NC QED-type covariant derivative. This second choice of SW map
differs from the first one by the gauge invariant term
$\theta^{ij}f_{i\mu}\star_2\partial_j\psi$, indicating a freedom in the choice of Seiberg-Witten map.
The second choice leads to the vertex
\begin{equation}
\Gamma^{\mu}_{}=iF(q,k)\left[\slashed k \tilde q^{\mu} + (q\theta
k)\gamma^{\mu} -\slashed q\tilde k^{\mu}\right].
\end{equation}
In general one can chose any superposition of the two SW maps $\hat\Phi_\text{alt}$ and $\hat\Phi_{}$,
but in this article we shall focus on the second choice as it is more natural from the point of view of gauge theory.

\section{Applications}

\subsection{Plasmon decay into $\bar\nu\nu$ pairs}

Our first phenomenological application of the new neutrino-photon vertex
(\ref{Feynrule}) will include a decay of transverse plasmon modes
into neutrino pairs,
which we then compare with the result obtained with perturbative methods (to first
order in $\theta$) in \cite{Schupp:2002up}.
Starting from the tree-level vector-like coupling to photons (\ref{Feynrule}),%
\footnote{A few parenthetical remarks are in order here.
In a different
model \cite{Ettefaghi:2007zz}, it is claimed that
Dirac masses for neutrinos are not consistent with NC gauge invariance
of the Yukawa terms. Our model features a vector-like
interaction, since both $\nu_L$ and $\nu_R$ are singlets under the residual
$U(1)_Q$ and
 Dirac mass terms for neutrinos are
allowed. The tree-level coupling to photons is experienced by both
neutrino chiralities even above the electroweak symmetry breaking scale.
Note also that the vertex (\ref{Feynrule}) is zero
for Majorana neutrinos, unless transition electromagnetic moments are
invoked. In all applications we will work with neutrinos with definite
chiralities, i.e.\ Weyl neutrinos, as the neutrino mass can be neglected.
The interaction (\ref{Feynrule}) is  generation-independent, and hereafter the
coupling constant is restored.}
standard $\gamma$-matrix techniques yield the amplitude squared for the process
$\gamma_{pl} \to
\bar\nu\nu$ summed over polarizations:
\begin{equation}
\left|M_{\rm NC}(\gamma_{pl} \to \bar\nu\nu)\right|^2
=4e^2(F(q,k))^2(q\theta k)^2\left(q^2+2m_\nu^2\right)\,. \label{M2}
\end{equation}
In the plasmon rest frame the total rate of decay into massless neutrinos involves the following phase
space integral over the outgoing neutrino momenta
\begin{equation}
\label{crosssection}
\begin{split}
&{\Gamma_{\rm NC}(\gamma_{pl} \to \bar\nu_{L \choose R} \nu_{L \choose R})}
=\frac{\alpha\,\omega_{pl}}{4\pi}
\int\limits_{0}^{\pi}\sin\vartheta
d\vartheta\int\limits_{0}^{2\pi}d\phi\sin^2 \frac{q\theta k}{2}\\&
=\frac{1}{4}\alpha \,\omega_{pl}\int\limits_{-1}^{1}dx
\bigg[1-(\cos Ax) J_0(B\sqrt{1-x^2})\bigg]\,,
\end{split}
\end{equation}
were $J_0$ is the zeroth order Bessel function of the first kind,
and\footnote{Here the dimensionless normalized matrix elements $c^{0i}$ are defined by
$\theta^{0i}=c^{0i}/\Lambda^2_{\rm NC}$, with $\sum\limits_{i=1}^3 |c^{0i}|^2=1$.}
\begin{equation}
A\equiv\frac{c_{03}\,\omega^2_{pl}}{2\Lambda^2_{\rm NC}},\quad
B\equiv\frac{\omega^2_{pl}}{2\Lambda^2_{\rm NC}}\sqrt{c^2_{01}+c^2_{02}}\,.
\label{AB}
\end{equation}
The plasma frequency $\omega_{pl}$ is defined as the frequency of plasmons
at $|\vec{q}| = 0$. In the  regime where the motion of background electrons
is irrelevant, i.e.\ $q_0 >2m_e$ and $|\vec{q}| > m_e$, the dispersion
relation for transverse and longitudinal waves can be calculated
analytically, giving (see e. g. \cite{Grasso:1993bp})
\begin{equation}
\omega^2_{pl} = {\cal R}e \,\Pi_{T}(q_{0}, |\vec{q}| = 0) = \frac{e^2 T^2}{9}\,,
\label{plasdisprel}
\end{equation}
where ${\cal R}e \,\Pi_{T}$ is the transverse part of the
one-loop contribution to the photon self-energy at
finite temperature/density and $T$ is the temperature.
The integral \eqref{crosssection} can be solved analytically (see Appendix):
\begin{equation}
{\Gamma_{\rm NC}(\gamma_{pl} \to \bar\nu_{L \choose R} \nu_{L \choose R})}
=\frac{\alpha}{2} \,\omega_{pl}
\left(1-\frac{\sin\xi}{\xi}\right)~,
\label{NCrate}
\end{equation}
where $\xi={\omega_{pl}^2}/{(2\Lambda_{\rm NC}^2)}$ and
$\alpha$ is the fine structure constant.

\begin{figure}[top]
\centerline{\includegraphics[width=8.5cm,angle=0]{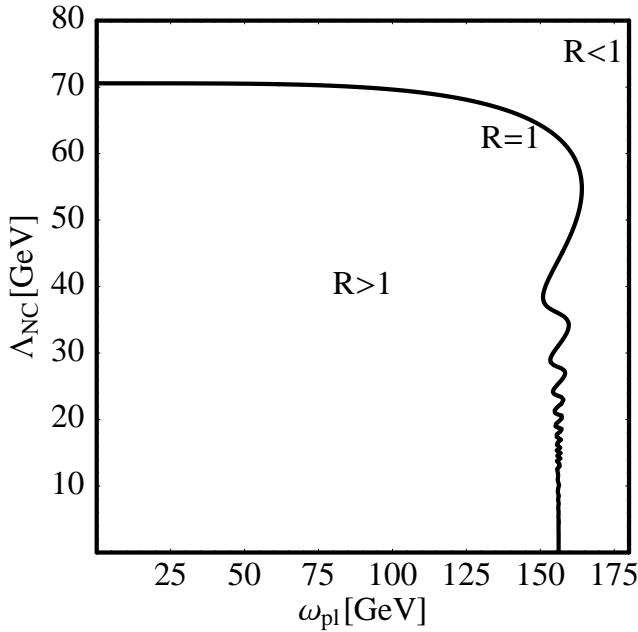}}
\caption{The plot of the scale of noncommutativity $\Lambda_{\rm NC}$ versus
the  plasmon frequency $\omega_{pl}$ with $R=1$. }
\label{fig:RvsOmegaLambda}
\end{figure}

The Standard Model (SM) neutrino-penguin-loop decay rate for transverse
plasmons (of energy $E_{\gamma}$)
into neutrinos is proportional
to $\omega_{pl}^6/E_{\gamma}$ \cite{Altherr:1993hb}.
Comparing the SM rate to our NC rate (\ref{NCrate}) for a plasmon at rest, and taking into account that
our NC photon-neutrino interaction (\ref{Feynrule})
has equal strength for both neutrino chiralities we obtain the ratio
\begin{equation}
\begin{split}
R\equiv&\frac{\sum_{\mbox{\rm\scriptsize flavors}}{\Gamma_{\rm NC}(\gamma_{pl}\to\bar\nu_L\nu_L+\bar\nu_R\nu_R)}}
{{\sum_{\mbox{\rm\scriptsize flavors}}\Gamma_{\rm SM}(\gamma_{pl} \to \bar\nu_L \nu_L)}}
\\=& \frac{3\cdot48 \pi^2\alpha^2}
{(c_{\nu_e}^2+c_{\nu_{\mu}}^2+c_{\nu_{\tau}}^2)\,G_F^2\,\omega_{pl}^4}
\left(1-\frac{\sin\xi}{\xi}\right)\,,
\label{nrate}
\end{split}
\end{equation}
see Figure~\ref{fig:RvsOmegaLambda}. For $\nu_e$, we have $c_{\rm v}=\frac{1}{2} + 2\sin^2\Theta_W$,
while for $\nu_\mu$ and $\nu_\tau$ we have $c_{\rm v}=-\frac{1}{2} + 2\sin^2\Theta_W$.
At small $\omega_{pl}$ and reasonably low NC scale $\Lambda_{\rm NC}$ the NC rate clearly dominates,
while for large $\omega_{pl}$ the Standard Model  rate dominates
regardless of the value of $\Lambda_{\rm NC}$.

Comparing to the previous result \cite{Schupp:2002up}, we note
an overall suppression factor $1-\frac{\sin\xi}{\xi}$ here, which,
depending on the parameters, may assume any value between 0 (for $\xi = 0$) and $1.22$ (for $\xi \approx 4.5$).
The previous result was perturbative in $\theta$, corresponding to an expansion of
the suppression factor as a power series of $\xi$
\begin{equation}
1-\frac{\sin\xi}{\xi}=\frac{1}{6}\xi^2-\frac{1}{120}\xi^4+\mathcal
O(\xi^6) \,.
\label{sinxi}
\end{equation}
The perturbative approach clearly fails, when $\xi$ is considerably larger than one,
while the new $\theta$-exact results remain valid.
Using
Eqs.~(\ref{NCrate}) and (\ref{sinxi}) for small $\xi$, we recover the old result \cite{Schupp:2002up},
thus showing consistency of the new computation.

Keeping just the first term in (\ref{sinxi}), the
ratio $R$ becomes independent of the plasma frequency $\omega_{pl}$. This
feature is reflected in the approximately constant $R=1$
contour for small $\omega_{pl}$ shown in Fig.~\ref{fig:RvsOmegaLambda} and
gives a lower bound $\Lambda_{\rm NC} \stackrel{>}{\sim} 70$ GeV,
in agreement with \cite{Schupp:2002up}, where a bound on
$\Lambda_{\rm NC}$ was derived from the requirement that NC contributions to plasmon
decay in stars should not go beyond the Standard Model predictions.
In practice, the plasmon frequency in stars,
$\omega_{pl} \simeq 10$ keV, is too low to see
the effect of the modified interaction (\ref{pf:int}). In the following, we shall explore other examples.

\subsection{Neutrino charge radius}


We make use of the full $\theta$-exact expression for the plasmon decay rate to recompute
also NC neutrino charge radii for both chiralities. Noting \cite{Minkowski:2003jg}
\begin{equation}
\Gamma(\gamma_{pl} \rightarrow {\bar\nu}_{\rm L}\nu_{\rm L})=
\frac{\alpha}{144}\frac{q^6}{E_{\gamma}}
\left|\langle r^2_{\nu}\rangle\right|^2\,
\label{Rateradii}
\end{equation}
and using
(\ref{NCrate})
we find that the $\theta$-exact NC induced neutrino charge radius  is given by
\begin{equation}
\left|\langle r^2_{\nu}\rangle\right|
=\lim_{\omega_{pl} \rightarrow 0}
\frac{6\sqrt2}{\omega_{pl}^2}\sqrt{1-\frac{\sin\xi}{\xi}}\,.
\label{NCradii}
\end{equation}
The limit $\omega_{pl} \to 0$ implies, through the dispersion
relation $q^2 = \omega_{pl}^2$, the familiar $q^2 \to 0$ limit
entering the definition of the charge radius. It is interesting to note
that the expansion in the plasma frequency coincides with an
expansion in $\theta$, because the effect of the full $\theta$-exact interaction enters only through the
parameter $\xi$. Therefore, the limit $\omega_{pl} \to 0$ picks up only the
first term in (\ref{sinxi}) and that corresponds to the
first-order-in-$\theta$ result, as stated before. This lucky coincidence implies
that there are no $\theta$-exact corrections to
the first-order-in-$\theta$ charge radius that was obtained earlier
 \cite{Minkowski:2003jg}:
\begin{equation}
|\langle r^2_{\nu} \rangle| = \frac{\sqrt{3}}{\Lambda_{\rm NC}^2} \;.
\label{radius}
\end{equation}
Note that the $\theta$-parameter -- when interpreted as  the length scale of
the fuzziness of spacetime which
arises as a consequence of space-space uncertainty relations -- directly runs
up a charge radius for a neutrino (by giving spatial extent to a point
particle), as expected.

With (\ref{radius}) at hand, one can immediately place a constraint on
$\Lambda_{\rm NC}$ by employing a very stringent bound on $\langle r^2_{\nu_R}\rangle$ based on
SN1987A \cite{Grifols:1989vi}.
With $\langle r^2_{\nu_R}\rangle \, \stackrel{<}{\sim} \, 2 \times 10^{-33} \rm cm^2$,
and using (\ref{radius}) one obtains $\Lambda_{\rm NC} \stackrel{>}{\sim} 0.6$ TeV.

\subsection{Big Bang Nucleosynthesis (BBN)}

Over the past decades, BBN has established itself as one of the most
powerful available probes of physics beyond the Standard Model, giving
many interesting constraints on particle properties (an extensive summary is
available, for instance, in \cite{Sarkar:1996}). One uses it to parametrize the energy
density of new relativistic particles
at the time of BBN in
terms of the effective number of additional neutrino species, $\Delta
N_{\nu}$, whose determination involves both a lower limit on
the barion-to-photon ratio ($\eta \equiv n_b/n_{\gamma }$) as well as an upper
bound
on the primordial mass fraction of $^{4}$He, $Y_p $ \cite{Olive:1980bu}. The energy
density of three light right-handed (RH) neutrinos produced by plasmon decay
during BBN is equivalent to the
effective number $\Delta N_{\nu}$ of additional doublet neutrinos
\begin{equation}
\Delta N_{\nu} = 3
\left (\frac{T_{\nu_{R}}}{T_{\nu_{L}}} \right )^4 \;,
\label{Ntemp}
\end{equation}
where $T_{\nu_{L}}$
is the temperature of the SM neutrinos, being the same as
that of photons down to $T \sim 1$ MeV. A better limit on $\Delta
N_{\nu}$ leads to a smaller value of $T_{\nu_R}$  and consequently a higher
decoupling temperature of the RH neutrinos. For $\Delta N_{\nu} = 1$, one finds
$T_{dec} > T_C $, where $T_C
$ is the critical temperature for the deconfinement restoration phase
transition, $T_C \sim 200$ MeV. If $\Delta N_{\nu} \simeq 0.2$, then
$T_{dec}$  would be
close to a critical temperature of the electroweak phase transition,
$T_{dec} \lesssim \; 300 $ GeV. Unfortunately, with the WMAP value
for $\eta $ \cite{Spergel:2003cb}, $Y_p $ was predicted to increase
\cite{Cyburt:2004cq}, having a
tendency to loosen the tight bounds on $\Delta N_{\nu}$ that existed before.

The RH neutrino is commonly considered to decouple at the temperature
$T_{dec}$ when the condition
\begin{equation}
{\Gamma(\gamma_{pl} \to \bar\nu_{R} \nu_{R})}\simeq H(T_{dec})
\label{decouple}
\end{equation}
is satisfied. The plasma frequency  in this case is given by
\begin{equation}
\omega_{pl}= \frac{eT_{dec}}{3}\,g^{ch}_{*}\,,
\label{disprel}
\end{equation}
while the Hubble expansion rate satisfies
\begin{equation}
H(T) \simeq 1.66 \,g_{*} \frac{T_{dec}^2}{M_{Pl}}
\label{HT}\;,
\end{equation}
where $g_{*}$ and $g^{ch}_{*}$ are the degrees of freedom specifying the
entropy of the interacting species for all and charged species,
respectively; $M_{Pl}$ is the Planck mass.

Computing the decoupling temperature $T_{dec}$ based on the assumption that
the decay rate (\ref{decouple}) is solely due to noncommutative effects and comparing
with lower bounds on $T_{dec}$ that can be inferred from observational data, we can determine
lower bounds on the scale of noncommutativity $\Lambda_{\rm NC}$.
Proceeding in this spirit, one finds that Big Bang nucleosynthesis provides the following relation between the decoupling
temperature $T_{dec}$ and the noncommutative scale $\Lambda_{\rm NC}$,
assuming that (\ref{decouple}) is fully due to noncommutative contributions:
\begin{equation}
T_{dec} \simeq \frac{M_{Pl}e^3 g^{ch}_*}{39.84\pi
g_*}\left(1-\frac{\sin\xi}{\xi}\right),\quad \xi=\frac{e^2
(g^{ch}_*)^2 T_{dec}^2}{18\Lambda_{\rm NC}^2}\,.
\label{Tdecxi}
\end{equation}
Let us consider the pre-factor
\begin{equation}
\frac{M_{Pl}e^3 g^{ch}_*}{39.84\pi g_*}
=\frac{\pi^{\frac{1}{2}}\alpha^{\frac{3}{2}}g^{ch}_*}{4.98g_*}M_{Pl},
\label{4.98}
\end{equation}
and taking into account $\alpha\simeq 137^{-1}$, $g^{ch}_*/g_*\simeq 1$, we
have
\begin{equation}
\frac{\pi^{\frac{1}{2}}\alpha^{\frac{3}{2}}g^{ch}_*}{4.98g_*}M_{Pl}\simeq
2.22\times 10^{-4}M_{Pl}.
\label{2.22}
\end{equation}
Since this factor is amplified by the Planck mass,
we can test the full interaction (\ref{pf:int}) only for $\Delta N_{\nu} \ll 1$.
Unfortunately, such a precision in observational data is not expected
to be reached anytime soon. Thus, to
account for decoupling temperatures associated with cosmological phase
transitions ($200\;\rm
MeV/300$ GeV), we have to require
$\left(1-\frac{\sin\xi}{\xi}\right)\ll 1$. This only occurs when
$\xi\to 0$, so that within this regime we can use the leading order term
in the Taylor expansion in $\xi$. This gives
\begin{equation}
\begin{split}
T_{dec}\simeq&
\frac{\pi^{\frac{1}{2}}\alpha^{\frac{3}{2}}g^{ch}_*M_{Pl}}{4.98
g_*}\cdot\frac{\xi^2}{6}
\\=&\frac{\pi^{\frac{1}{2}}\alpha^{\frac{3}{2}}g^{ch}_*M_{Pl}}{4.98
g_*}\cdot\frac{4\pi\alpha
(g^{ch}_*)^2}{108}\cdot\left(\frac{T_{dec}}{\Lambda_{\rm NC}}\right)^4\,,
\end{split}
\label{T_dec}
\end{equation}
so that
\begin{equation}
\Lambda_{\rm
NC}\simeq\left(\frac{\pi^{\frac{3}{2}}\alpha^{\frac{5}{2}}
(g^{ch}_*)^3M_{Pl}T_{dec}^3}{4.98\cdot 27 g_*}\right)^{\frac{1}{4}}\,.
\label{LaNC}
\end{equation}
Now setting further $g_*\sim
g^{ch}_*\sim 100$, $M_{Pl}=1.22\times 10^{19}$ GeV and
$T_{dec}>200$ MeV (quark-hadron phase transition), a lower bound on $\Lambda_{\rm
NC}$ can be obtained as
\begin{equation}
\Lambda_{\rm NC}>3.68\; \rm TeV.
\label{8.35}
\end{equation}
For $T_{dec}>300$ GeV (eletro\-weak phase transition), we have
\begin{equation}
\Lambda_{\rm NC}>887 \; \rm TeV,
\label{10^3}
\end{equation}
confirming the previous results obtained by making use of scattering processes
of $\nu_R$s \cite{Horvat:2009cm}.

\begin{figure}[top]
\begin{center}
\includegraphics[width=8.5cm,angle=0]{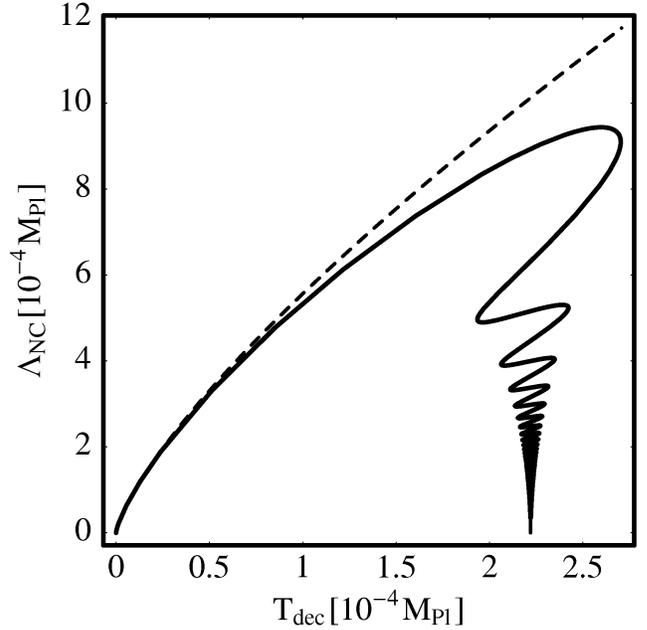}
\end{center}
\caption{The plot of the scale of noncommutativity $\Lambda_{\rm NC}$ versus
decoupling temperature $T_{dec}$. The dashed/full curve corresponds to the
perturbative/full solution, as given by
Eqs.~(\protect\ref{LaNC}) and (\protect\ref{Tdecxi}), respectively. In both curves we
set, for illustration purposes, $g_{*} = g_{*}^{ch} = 100$, and the logarithmic scaling of the fine
structure constant with temperature is ignored. The full curve reveals
$T_{dec}^{max} = 2.7 \times 10^{-4} M_{Pl}$ and $\Lambda_{\rm NC}^{max} =
9.4 \times 10^{-4} M_{Pl}$.}
\label{fig:LambdaVsTc}
\end{figure}

For the sake of demonstration, we have also studied (\ref{Tdecxi}) numerically, to
investigate how the plasmon rate obtained in the full theory affects determination
of $\Lambda_{\rm NC}$ when $T_{dec} \simeq 10^{-4} M_{Pl}$. The relation
(\ref{Tdecxi}) is depicted in Fig. \ref{fig:LambdaVsTc}, in which the
approximate result (\ref{LaNC}) is also superimposed for comparison.

Looking at the asymptotic behavior of the approximate (\ref{LaNC}) and exact (\ref{Tdecxi}) relations
between $\Lambda_{\rm NC}$ and $T_{dec}$, one notes
that Fig.~2 reveals quite a different behavior for them: While
the solution of (\ref{Tdecxi}), obtained by employing the leading order term in
$\xi$, shows no restriction on $T_{dec}$ all the way up till the beginning
of the radiation era, the solution of (\ref{LaNC}) reveals a maximal decoupling
temperature in the said epoch. (Of course, if inflation occurred
well below $T_{dec}^{max} \simeq 3 \times 10^{-4} M_{Pl}$, then
any distinction between the two solutions would practically disappear.) This
feature is accompanied by a maximal scale of noncommutativity, above which
the RH neutrinos can no longer retain thermal contact with the rest of the
universe. This is in contrast with the exact relation (\ref{Tdecxi}), where thermal
equilibrium is maintained for much larger $\Lambda_{\rm NC}$'s  (if inflation
occurred well above $T_{dec}^{max}$), and stops when $T_{dec}$
hits the reheating temperature. This means that the effect of the full
interaction (\ref{Feynrule}) is to bring about the maximal upper limit $\Lambda_{\rm
NC}^{max}
\simeq 9 \times 10^{-4} M_{Pl}$, occurring at a decoupling temperature
slightly below $T_{dec}^{max}$, which can be extracted by our
method.  Note that for decoupling temperatures lying in a region where the
oscillatory patterns inherent in (\ref{LaNC}) becomes manifest, it may seem
troublesome to infer a bound on the scale of noncommutativity
since several (many) $\Lambda_{\rm NC}$'s correspond to the same $T_{dec}$. In
these cases one chooses the highest $\Lambda_{\rm NC}$, otherwise
$\Lambda_{\rm NC}$'s
obtained with a much smaller decoupling temperature (where the oscillatory
term is shut down) would give much better lower limits. Again, this
characteristic is missing in the perturbative solution, where
better and better limits
on the NC scale are always accompanied  with  progressively  increasing
decoupling temperatures.

Note that these ranges of $T_{dec}$
are of course tremendously above the bounds that can be inferred from
current observational data.


\subsection{Ultrahigh Energy (UHE) Cosmic Rays}

The non-observation of UHE neutrino induced events in neutrino observatories
implies a strong model-independent constraint on the inelastic
neutrino-nucleon cross section \cite{Anchordoqui:2004ma}, which consequently gives a
constraint on the scale of noncommutativity for a NC gauge-field theory in which
neutrinos couple directly to photons \cite{Horvat:2010sr}. It was observed
\cite{Horvat:2010sr} that at energies as high as $10^{11}$ GeV, the usual expansion in $\theta$
is no longer meaningful. In order to fix the breakdown in
the perturbative expansion, a resummation in the neutrino-photon
vertex was undertaken \cite{Horvat:2010sr}. Although
devoid of a firm theoretical background and with only the zeroth order in
the Seiberg-Witten map employed, this {\sl ad hoc} approach did produce the correct $\sin$
term that we have now obtained, redoing the computation using the $\theta$-exact vertex  (\ref{Feynrule}).

Curiously enough, (\ref{Feynrule}) can be put in a much simpler form
if the NC vertex connects external (on-shell) neutrino lines. Indeed, with
the aid of the Dirac equation for free fields, the first term in (\ref{Feynrule})
vanishes if one line is on-shell, while the third term in (\ref{Feynrule})
vanishes if both lines are on-shell. Thus, for tree-level processes with
no internal neutrino lines, (\ref{Feynrule}) reduces to
\begin{equation}
\Gamma^{\mu} = 2 i\,e \,\gamma^{\mu} \;{\rm sin} \left (\frac{q \theta k}{2} \right )\,.
\label{adhocFR}
\end{equation}
This exactly coincides with the result \cite{Horvat:2010sr} obtained with
the {\sl ad-hoc} method. This way, the powerful bounds on $\Lambda_{\rm NC}$
obtained there, in the range 200-900 TeV (depending on a model for the
cosmogenic neutrino flux), get further credence as far as the
underlying theoretical background is concerned. Note, though, that
(\ref{adhocFR}) should not be used  to calculate
tree-level processes with internal neutrino lines, nor in calculations
involving loops. Hence, if one is, for instance, to
study the UV/IR mixing in the neutrino sector,
then the complete expression as given by (\ref{Feynrule}) should be used.

\section{Conclusion}

In summary, we showed that the tree-level tri-particle
decay $\gamma_{pl}\to\nu\bar\nu$ in the covariant noncommutative quantum gauge
theory based on Seiberg-Witten maps can be computed without an
expansion over the noncommutative parameter $\theta$.
As an application, we focus on plasmon decay into neutrinos,
reconsidering previous computations that were done with less sophisticated tools and deriving new
bounds on the scale of noncommutativity.

Comparing to
previous results, the total decay rate is modified by a factor
which remains finite throughout all energy scales. Thus the new results
behave much better than the $\theta$-expansion method when ultra
high energy processes are considered. We expect that similar
control on the high energy behavior can be extended to $\theta$-exact perturbation theory involving
more than three external fields in the near future. This would provide a considerably
improved theoretical basis for research work in the field of noncommutative
particle phenomenology.

\noindent
\acknowledgments
J.T. would like to acknowledege support of W. Hollik, and MPI Munich for hospitality.
The work of R.H., D.K and J.T. are supported by
the Croatian Ministry of Science, Education and Sports
under Contracts Nos. 0098-0982930-2872 and 0098-0982930-2900, respectively.
The work of J.Y. was supported by the Croatian NSF and the IRB Zagreb,
and by the German Research Foundation (Deutsche
Forschungsgemeinschaft (DFG)) through the Institutional Strategy of the
University of G\"ottingen.

\appendix
\section{Integral}

\noindent
We evaluate the integral in (\ref{crosssection})
\begin{equation}
\begin{split}
\mathcal I=&\int\limits_{0}^{1}dx (\cos Ax) J_0(B\sqrt{1-x^2})
\\=&
\int\limits^1_0 dx \cos
(A\sqrt{1-x^2})\frac{xJ_0(Bx)}{\sqrt{1-x^2}}
\end{split}
\end{equation}
analytically by Taylor expansion of both $\cos x$ and $J_0(x)$
\begin{equation}
\cos x=\sum\limits^\infty_{k=0} \frac{(-)^k x^{2k}}{(2k)!}\,,
\hspace{1cm}
J_0(x)=\sum\limits^\infty_{k=0}\frac{(-)^k}{(k!)^2}\left(\frac{x}{2}\right)^{2k}\,.
\end{equation}
We consider the integral over the product of the $m$-th order
term in the Taylor expansion of $\cos (A\sqrt{1-x^2})$ and the $n$-th
order term in the Taylor expansion of $J_0(Bx)$
\begin{equation}
\begin{split}
\mathcal I (m,n)=&\int\limits^1_0 dx\frac{(-)^m
(A\sqrt{1-x^2})^{2m}}{(2m)!}\frac{(-)^nx}{(n!)^2\sqrt{1-x^2}}\left(\frac{Bx}{2}\right)^{2n}
\\
=&\frac{(-)^{m+n}(A^2)^m(B^2)^n}{(2m)!(n!)^22^{2n+1}}\int\limits^1_0
dx(1-x)^{\frac{2m-1}{2}}x^{n}.
\end{split}
\end{equation}
The integrals over $x$ can now be done by successive integration by parts,
yielding
\begin{eqnarray}
\int\limits^1_0
dx(1-x)^{\frac{2m-1}{2}}x^{n}&=&\frac{n!2^{n+1}}{(2m+1)(2m+3)...(2m+2n+1)}
\nonumber \\
&=&\frac{n!2^{2n+1}(2m)!(n+m)!}{(2(m+n)+1)!m!}
\end{eqnarray}
and
\begin{equation}
\mathcal I
(m,n)=\frac{(-)^{m+n}}{(2(m+n)+1)!}\frac{(n+m)!}{n!m!}(A^2)^m(B^2)^n.
\end{equation}
Re-summing
\begin{equation}
\sum\limits_{n+m=l}\mathcal I(m,n)=\frac{(-)^l}{(2l+1)!}\left(
\sqrt{A^2+B^2}\right)^{2l}
\end{equation}
and noting
\begin{equation}
\sin x=\sum\limits_{l=0}^\infty \frac{(-)^l
x^{2l+1}}{(2l+1)!}=x\sum\limits_{l=0}^\infty \frac{(-)^l
x^{2l}}{(2l+1)!}\,,
\end{equation}
we finally obtain the surprisingly simple result
\begin{equation}
\mathcal I=\frac{\sin \sqrt{A^2+B^2}}{\sqrt{A^2+B^2}} \,
\equiv \frac{\sin \xi}{\xi}\,. 
\end{equation}

\end{document}